\documentclass{article} \usepackage{slashed,epsfig,amsmath,amssymb,enumitem} \usepackage[papersize={8.5in,11in}]{geometry}
   \usepackage{bm}
\usepackage{graphicx}
\newcommand{\be}{\begin{equation}}
\newcommand{\ee}{\end{equation}}
\title{Stochastic Fluctuations and Brownian Motion Detection of Gravitons}
\author{J. W. Moffat\\
Perimeter Institute for Theoretical Physics, Waterloo, Ontario N2L 2Y5, Canada\\
and\\
Department of Physics and Astronomy, University of Waterloo, Waterloo,\\
Ontario N2L 3G1, Canada}
\begin{document}
\maketitle


\begin{abstract}
We propose a way to detect gravitons by replicating the Brownian motion experiment. The number $N_g$ of gravitons can be large enough for the stochastic gravitational noise produced by them to displace a massive test particle in a physical system, allowing for the detection of gravitons. Possible experiments to detect gravitons are proposed involving collective stochastic fluctuations due to a large number of gravitons, causing a Brownian motion displacement of a massive test body. Gravitational wave experiments involving advanced interferometer techniques and mirrors could detect the large collective number of gravitons, and could detect Brownian motion of test particles in the detectors' component mirrors. The problem of reducing thermal and other background noise is investigated.
\end{abstract}

\maketitle

\section{Introduction}

An important problem in modern physics is to understand the role of quantum physics in gravity. A quantum gravity based on a quantization of the spacetime manifold geometry can be formulated employing the methods of effective quantum field theory assuming that the quantum particle is a graviton. The problem that has beset this approach has been the lack of experimental data allowing for a test of any quantum gravity predictions. At the basic level the length and time scales associated with Newton's gravitational constant $G_N$ are $l_p=\sqrt{G_N\hbar/c^3}\approx 10^{-35} m$ and $t_p=l_p/c\approx 10^{-44}s$. In effective field theory (EFT) and quantum mechanics, these scales are far beyond the reach of observational experiments, when single or a few gravitons are describing calculations of graviton scattering amplitudes and cross sections~\cite{Boughn,Dyson}.

In refs.~\cite{Moffat1,Moffat2}, gravity was treated as a stochastic phenomenon, based on fluctuation of the metric tensor of general relativity. By using a $(3+1)$ slicing of spacetime, a covariant Langevin equation of motion of a test particle was derived, generalizing the classical geodesic motion of a massive test particle. A covariant Langevin equation for the dynamical conjugate momentum and a Fokker-Planck equation for its probability distribution was derived.

Several proposals have been made for experiments to detect gravitons without resorting to a physically unrealistic large mass detector or accelerator~\cite{Pikovski1,Pikovski2,Tabletop,Parikh1,Parikh2,Parikh3,Cho,Kanno,Haba1,Haba2,Zurek1,Zurek2,Lee1,Lee2,Badurina,Bub,Fuentes,HowlFuentes,Carney}. A proposal is to attempt to quantum entangle two masses through the influence of gravity. Such experimental attempts face the orders of magnitude of the size of test masses that can be influenced by gravity and the possible quantum gravity entanglement of test masses. In the following, we will propose detecting gravitons by a repeat of the stochastic Brownian motion experiments that detected atoms and molecules~\cite{Brown,Einstein,Perrin}.

In ref.~\cite{Moffatgraviton}, the gravitational wave strain $\Delta L/L\sim 10^{-21}$ was predicted from the coherent classical gravitational wave and quantum graviton calculations for a large number of gravitons.

\section{Gravitational Brownian motion detection of gravitons}

We approach experimental detection of gravitons as follows. We consider the displacement $\Delta x$ of a massive test particle caused by stochastic Brownian fluctuations along with interaction with gravitons.

The weak gravitational field is described by
\be
g_{\mu\nu}=\eta_{\mu\nu}+\kappa h_{\mu\nu},
\ee
where $\eta_{\mu\nu}$ is the Minkowski metric tensor and $\kappa=\sqrt{8\pi G_N/c^4}$. The coupling between matter and the gravitational field is proportional to the stress-energy momentum tensor $T_{\mu\nu}$ and the gravitational field perturbation $h_{\mu\nu}$. The displacement of the massive test particle depends on the strength of this coupling and the amplitude of the stochastic perturbations caused by the gravitons.

We postulate that the gravitational displacement $\Delta x_g$ is given by
\be
\Delta x_g=\frac{G_N}{c^2}m_tN_gf(\rho,\eta ...),
\ee
where $f(\rho,\eta,...)$ is a dimensionless function of the density $\rho$ and viscosity $\eta$ of a physical system. For a sufficiently large number of gravitons, $N_g$, the displacement of the massive test particle, in addition to its stochastic Brownian motion fluctuations could allow for a detection of the gravitons. The stochastic Brownian motion displacement in the medium $\Delta x_m$ will vary as the square root of the time,
\be
\Delta x_m=\bigg[\bigg(\frac{k_BT}{6\pi\eta R}\bigg)t\bigg]^{1/2},
\ee
where $R$ is the size of the test particle~\cite{Einstein}. The Brownian motion of the test particle can be measured as $v_t=\Delta x_m/\Delta t$, when $\Delta t << \tau$, where $\tau$ is the momentum relaxation time. The velocity data can be verified by the Maxwell-Boltzmann velocity distribution, and the equipartition theorem for a Brownian test particle.

The total displacement of the test particle, assuming its stochastic Brownian motion and the influence due to the graviton flux are statistically independent, would be given by the RMS sum:
\be
\Delta x = \sqrt{\Delta x_g^2 + \Delta x_m^2}.
\ee

\section{Background thermal noise and graviton signal}

To distinguish the displacement $\Delta x$ of a massive test particle caused by graviton fluctuations from the background thermal noise in a Brownian motion experiment, we can consider the following approaches:

1) Temperature dependence: The thermal noise due to the fluid or gas atoms and molecules follows a temperature-dependent distribution, such as the Maxwell-Boltzmann distribution. In contrast, the graviton-induced displacement is expected to be independent of temperature. By conducting the experiment at different temperatures and comparing the observed displacements, $\Delta x$, we may be able to separate the graviton-induced effect from the thermal noise.

2) Frequency analysis: The stochastic fluctuations caused by gravitons may have a different frequency spectrum compared to the thermal noise. By analyzing the frequency components of the observed displacement, $\Delta x$, using techniques like Fourier analysis, we might be able to identify the graviton-induced signal, if it has a distinct frequency signature.

3) Particle mass dependence: The displacement $\Delta x$ caused by gravitons should depend on the mass $m_t$ of the test particle, as the gravitational interaction is proportional to mass. By using test particles of different masses and comparing their displacement, we can check if the observed effect scales with mass in a way consistent with gravitational interactions.

4) Shielding and isolation: To minimize the background thermal noise, the experiment can be conducted in a well-isolated and shielded environment, such as a high-vacuum chamber with effective thermal insulation. This can help reduce the thermal noise and make the graviton-induced signal more prominent.

5) Statistical analysis: By collecting a large number of measurements and performing statistical analysis, such as calculating the variance of high-order moments of the displacement $\Delta x$ distribution, we may be able to distinguish the graviton-induced effect from the thermal noise. The graviton-induced displacement $\Delta x$ is expected to have a different statistical distribution compared to the thermal noise.

Let us consider approach (5). We assume that the the displacement $\Delta x$ is measured at regular time intervals $\Delta t$. We can then calculate the moments of the displacement distribution and compare the two cases. For the thermal noise distribution: $\langle\Delta x\rangle=0$ and $\langle (\Delta x)^2\rangle=2D\Delta t$, where $D$ is the diffusion coefficient, which depends on the temperature and the properties of the particle in the medium.

In the presence of gravitons, the displacement $\Delta x$ is expected to deviate from the normal distribution due to the additional gravitational action. The exact form of the graviton-induced distribution depends on the specific theoretical model, but it is generally expected to have non-zero higher-order moments, such as skewness and kurtosis.

For example, if the graviton-induced displacement has a non-zero third moment (skewness), it would indicate an asymmetry in the distribution: $\langle(\Delta x)^3\rangle\neq 0$. Similarly, a non-zero fourth-order moment (kurtosis) would indicate a deviation from the Gaussian shape of the thermal noise distribution:
$\langle(\Delta x)^4\rangle\neq 3(\langle(\Delta x)^2\rangle)^2$.

To distinguish between the two distributions, one would need to perform a large number of measurements
$\Delta x$ at regular time intervals $\Delta t$. A calculation would be performed of the displacement distribution, such as the variance, skewness and kurtosis and the results compared to the calculate moments with the expected values for the thermal noise expectations. If the calculated moments deviate significantly from the thermal noise expectations, it could indicate the presence of a graviton induced effect. Statistical tests could be performed, such as the chi-squared test or the Kolmogorov-Smirnov test, to quantify the significance of the deviation from the thermal noise distribution.

\section{Graviton detection experiments}

It is important that detecting gravitons through this method would require extremely precise measurements and a large number of data points, as the gravitational interaction is expected to be very weak compared to the thermal noise. Additionally, other sources of noise and systematic errors would need to be carefully controlled and accounted for in the analysis.

In a hypothetical experiment to detect gravitons, using a setup analogous to a Brownian motion experiment, the goal would be to create a system where a large coherent collection of gravitons interacts with massive particles, causing them to exhibit random motion similar to that observed in traditional Brownian motion experiments. Such an experiment would require significant advancements in our understanding of gravitons and our ability to manipulate them.

A source capable of producing a large, coherent collection of gravitons would be required. This is a significant challenge. In a traditional Brownian motion experiment, the fluid consists of atoms or molecules that constantly collide with the larger test particles, causing them to exhibit random motion. In the graviton experiment we would need to replace the fluid with a medium that allows gravitons to propagate and interact coherently with the larger test particles. The massive test particles would need to be influenced by gravitational interactions. They should have a large enough mass to be influenced by the gravitons, but small enough to exhibit observable random motion.

Observing the motion of the test particles would require a highly sensitive detection system capable of tracking the positions of the test particles with great precision. This could involve advanced interferometer techniques and mirrors, such as those used in the gravitational wave detectors Ligo/Virgo, or novel methods based on the interaction between the test particle and electromagnetic fields~\cite{Pikovski1,Pikovski2,Parikh1,Parikh2,Parikh3,Cho,Kanno,Haba1,Haba2,Zurek1,Carney}.

Let us consider the detection of gravitational waves at the LIGO/Virgo observatories~\cite{Abbott}. The energy scale at which the gravitational interaction becomes comparable to the other fundamental forces is the Planck scale, which is characterized by the Planck energy, $E_p=1.22\times 10^{19}$ GeV, Planck mass $m_p=2.18\times 10^{-8}$ kg and Planck frequency $f_p=2.93\times 10^{43}$ Hz. It is clear that to achieve the Planck frequency in a gravitational wave detector is far beyond our current technological capabilities, making the detection of a single graviton not feasible.

It would be possible to take advantage of the large number of gravitons in a gravitational wave, making the coherent collection of the gravitons detectable more feasible, instead of the detection of a single graviton.
The gravitational wave would be considered a coherent state, similar to how a laser is a coherent state of many photons. In a coherent state, the gravitons are in phase and act together, enhancing their collective effect.

Coherence of gravitons in black hole mergers: During the inspiral and merger of two black holes, the gravitational field becomes extremely strong and dynamical. The resulting gravitational waves can be described as a coherent state of many gravitons, all in phase and acting together. This coherence is maintained as the gravitational wave propagates through space. In a laser, the photons are coherent and in phase, which allows them to constructively interfere and produce a strong, focused beam of light. Similarly, the coherent gravitons in a gravitational wave from a black hole merger constructively interfere, resulting in a strong, detectable signal.

The number of gravitons contained in a gravitational wave detection depends on the strength of the gravitational wave and its frequency. To estimate the number of gravitons in a typical gravitational wave detection, we can use the example of the first direct detection of gravitational waves by LIGO in 2015, known as GW150914~\cite{Abbott}. The gravitational wave strain of GW150914 was approximately $h\sim 10^{-21}$. The peak frequency f of the gravitational wave was around $f\sim 250$ Hz. The total energy $E_{GW}$ radiated in the form of gravitational waves was about 3 solar masses, or approximately $E_{GW}\sim 3.38\times 10^{66}$ eV.

let us calculate the energy $E_g$ of a single graviton at the peak frequency $f$ using the Planck-Einstein relation, $E = hf$, and the frequency $f\sim 250$ Hz. We get $E_g=1.04\times 10^{-10}$ eV. We can now estimate the number of gravitons in the gravitational wave by dividing the total energy $E_{GW}$, radiated in the GW150914 event by the energy of a single graviton $E_g$, giving $N_g\sim 10^{78}$.

This is a large number of gravitons, demonstrating the collective strength of the gravitational wave signal. It is important to note that this is an estimate based on the assumption that all the gravitons have the same frequency, which is a simplification. In reality, the gravitational wave has a range of frequencies, and the number of gravitons at each frequency would vary.

A typical gravitational wave detection, like GW150914, can contain an extremely large number of gravitons on the order of $10^{78}$ or more. This collective effect of many coherent gravitons is what allows gravitational wave detectors to measure the tiny distortions in spacetime caused by passing gravitational waves, even though the interaction of individual gravitons with matter is exceptionally weak~\cite{Moffatgraviton}.

Let us consider using a Brownian motion type of experimental detection to determine the existence of a large collective number of gravitons in a gravitational wave. While this concept has not been directly implemented in current gravitational wave detectors like LIGO and Virgo, it is worth exploring as a potential avenue for future research.

Brownian motion is the random motion of particles suspended in a fluid, caused by collisions with the molecules of the fluid. In the context of gravitational wave detection, we can consider the idea to look for a random Brownian motion of the detector's components (such as the mirrors) caused by collisions with the gravitons in the gravitational wave.

For this approach to work, the detector's components would need to be extremely sensitive to the tiny momentum transfers from graviton collisions. This would require a significant improvement in the sensitivity of current gravitational wave detectors, as well as a reduction in other sources of noise that could mask the effect of graviton collisions.

If the detector's components could be made sensitive enough to detect individual graviton collisions, the next step would be to measure the collective effect of the large number of gravitons in the gravitational wave. This could potentially be done by analyzing the statistical properties of the random motion caused by the graviton collisions, such as the power spectrum or correlation functions.

A major challenge in this approach would be distinguishing the random motion caused by graviton collisions from other sources of noise, such as thermal noise, seismic noise, or quantum fluctuations in the detector's components. This would require a deep understanding of the detector's noise sources and the development of sophisticated data analysis techniques.

Implementing this idea would require significant advances in both theoretical and experimental physics. On the theoretical side, a better understanding of how gravitons interact with matter at the quantum level would be necessary. On the experimental side, the development of ultra-sensitive detectors and noise reduction techniques would be crucial.

\section{Conclusions}

It is essential to explore ideas to push the boundaries of our understanding of quantum gravity and to develop new experimental techniques to detect the graviton and confirm that it is the basic quantum particle of gravitation. To possibly detect gravitons in a gravitational wave signal would require performing gravitational wave experiments with a special design of interferometer arms and mirrors that can overcome the problem of background noise and allow for the detection of graviton noise.

It is possible that a stochastic Brownian motion experiment could lead to the detection of gravitons. Several experimental techniques can be followed to isolate the spacetime graviton stochastic fluctuations and graviton noise from background thermal fluctuations and noise of atoms and molecules in a physical medium. The weakness of gravity compared to the other forces of nature may be overcome by experiments concentrating on the statistical nature of the graviton as a many particle phenomenon.

Designing and conducting experiments to detect gravitons is challenging, requiring significant theoretical and technological advancements. Key areas of research that could contribute to this goal include quantum gravity theories that provide a better understanding of gravitons and their properties. Development of novel materials or substances that could act as a medium for graviton propagation. An improvement of ultra-sensitive detection methods for tracking the motion of particles at the microscopic level.

The experimental detection of gravitons presupposes the traditional understanding of quantum gravity as a quantized gravitational theory in four-dimensional spacetime, with a spin-2 graviton, similar to spin-1 photons in quantum electrodynamics. This solution can only succeed if an experiment can be performed to detect gravitons and allow for verification of the methodology of quantizing gravity. As our understanding of quantum gravity and the properties of gravitons evolves, the feasibility and design of a graviton detection experiment may significantly change in the future.

\section*{Acknowledgments}

I thank Laurent Friedel, Martin Green and Viktor Toth for helpful discussions. Research at the Perimeter Institute for Theoretical Physics is supported by the Government of Canada through industry Canada and by the Province of Ontario through the Ministry of Research and Innovation (MRI).


\begin{thebibliography}{99}

\bibitem{Boughn} S. Broughn and T. Rothman, Class. Quant. Grav., 23, 5839 (2006).

\bibitem{Dyson} F. J. Dyson, Poincar\'e Prize Lecture, XVIIth International Congress on Mathematical Physics, Aalborg, Denmark, August 6, 2012, World Scientific, 670 (2013).

\bibitem{Moffat1} J. W. Moffat, Physical Review, D56, 6264 (1977), arxiv: gr-qc/9610067.

\bibitem{Moffat2} J. W. Moffat, Proceedings of the workshop: Very High Energy Phenomena in the Universe,
XXXIInd Rencontres de Moriond, Les Arcs, France, 1997, arxiv:gr-qc/9703032.

\bibitem{Pikovski1} G. Tobar et al., Nature Communications 15, 7229 (2024), arxiv:astro-ph/2308.15440.

\bibitem{Pikovski2} V. Shenderov et al., arxiv:gr-qc/2407.11929.

\bibitem{Tabletop} D. Carney, P. C. E. Stamp, and J. M. Taylor, Class. Quant. Grav. 36, 3, 2019, arxiv:quant-ph/1807.11494.

\bibitem{Parikh1} M. Parikh, F. Wilczek and G. Zahariade, Physical Review, D104, 046021 (2021), arxiv:2010.08208.

\bibitem{Parikh2} M. Parikh, F. Wilczek and G. Zahariade, Physical Review Letters, 127, 081602 (2021), arxiv:2010.08205.

\bibitem{Parikh3} M. Parikh, F. Wilczek and G. Zahariade, International Journal of Modern Physics,  D29, 2042001 (2020), arxiv:2005.07211.

\bibitem{Cho} H. T. Cho and B. L. Hu, Physical Review, D105, 086004 (2022), arxiv:2112.08174.

\bibitem{Kanno} K. Kanno, J. Soda and J. Tokuda, Physical Review, D103, 044017 (2021), arxiv:2007.09838.

\bibitem{Haba1} Z. Haba, Universe, 7, 117 (2021), arxiv:2104.10433.

\bibitem{Haba2} Z. Haba, International Journal of Modern Physics, D32, 2350005 (2023), arxiv:2202.06125.

\bibitem{Zurek1} K. M. Zurek, Physics Letters B, 826, 136910 (2022), arxiv:hep-th/2012.05870.

\bibitem{Zurek2} K. M. Zurek, arxiv:[gr-qc]/2012.05870.

\bibitem{Lee1} V. S. H. Lee, K. M. Zurek and Y. Chen, Physical Review, D109, 8, 084005 (2024), axiv:[gr-qc]/2312.06757.

\bibitem{Lee2} V. S. H. Lee and K. M. Zurek, arxi:[gr-qc]/2408.03828.

\bibitem{Badurina} L. Badurina et al., arxiv:[gr-qc]/2409.03828.

\bibitem{Bub} M. W. Bub et al., arxiv:[gr-qc]/2305.11224.

\bibitem{Fuentes} R. Howl, I. Fuentes, and R. Penrose, New Journal of Physics, 21, 043047 (2019), arxiv:quant-phys/1812.04630.

\bibitem{HowlFuentes} R. Howl and I. Fuentes, arxiv:quant-phys/2103.02618.

\bibitem{Carney} D. Carney, V. Domcke, and N. L. Rodd, Physical Review, D109, 044009 (2024), arxiv:hep-th/2308.12988.

\bibitem{Brown} R. Brown, Royal Society, London, vol. 1, 1866.

\bibitem{Einstein} A. Einstein, Annalen der Physik, 322, 549 1905.

\bibitem{Perrin} J. Perrin, Annnales de Chimie et de Physique, 18, 5-114 (1909).

\bibitem{Moffatgraviton} J. W. Moffat, arxiv:gr-qc/2411.06265.

\bibitem{Abbott} P. Abbott et al. (LIGO Scientific Collaboration and Virgo Collaboration) Physical Review Letters, 116, 061102 (2016); Physical Review D95, 04003 (2017, arxiv: astro-ph.HE/1611.02972.

\end{thebibliography}
\end{document}